\documentclass[conference]{IEEEtran}
\IEEEoverridecommandlockouts
\UseRawInputEncoding

\usepackage{cite}
\usepackage{amsmath,amssymb,amsfonts}
\usepackage{algorithmic}
\usepackage{graphicx}
\usepackage{textcomp}
\usepackage{xcolor}
\usepackage{xspace}

\def\BibTeX{{\rm B\kern-.05em{\sc i\kern-.025em b}\kern-.08em T\kern-.1667em\lower.7ex\hbox{E}\kern-.125emX}}

\usepackage{pgfplots}
\def\BibTeX{{\rm B\kern-.05em{\sc i\kern-.025em b}\kern-.08em
    T\kern-.1667em\lower.7ex\hbox{E}\kern-.125emX}}
\begin{document}

\newcommand{\kiglis}{{\sc Kiglis}\xspace}


\title{\kiglis: Smart Networks for Smart Cities\\
\thanks{This work results from the \kiglis project supported by the German Federal Ministry of Education and Research (BMBF), grant numbers 16KIS1231, 16KIS1227K, 16KIS1228,  16KIS1230, 16KIS1232 and 16KIS1233.}
}

\author{\IEEEauthorblockN{Daniel Bogdoll\IEEEauthorrefmark{1},
Patrick Matalla\IEEEauthorrefmark{3},
Christoph F\"ullner\IEEEauthorrefmark{3},
Christian Raack\IEEEauthorrefmark{4},
Shi Li\IEEEauthorrefmark{6},
Tobias K\"afer\IEEEauthorrefmark{5},\\
Stefan Orf\IEEEauthorrefmark{1},
Marc Ren\'e Zofka\IEEEauthorrefmark{1},
Finn Sartoris\IEEEauthorrefmark{1},
Christoph Schweikert\IEEEauthorrefmark{7},
Thomas Pfeiffer\IEEEauthorrefmark{2},\\
Andr\'e Richter\IEEEauthorrefmark{6},
Sebastian Randel\IEEEauthorrefmark{3},
Rene Bonk\IEEEauthorrefmark{2}}

\IEEEauthorblockA{\IEEEauthorrefmark{1}FZI Research Center for Information Technology, Germany.
Email: bogdoll@fzi.de}
\IEEEauthorblockA{\IEEEauthorrefmark{2}Nokia Bell Labs, Germany.
Email: rene.bonk@nokia-bell-labs.com}
\IEEEauthorblockA{\IEEEauthorrefmark{3}IPQ at Karlsruhe Institute of Technology, Germany. 
Email: sebastian.randel@kit.edu}
\IEEEauthorblockA{\IEEEauthorrefmark{4}atesio GmbH, Germany.
Email: raack@atesio.de}
\IEEEauthorblockA{\IEEEauthorrefmark{5}AIFB at Karlsruhe Institute of Technology, Germany.
Email: tobias.kaefer@kit.edu}
\IEEEauthorblockA{\IEEEauthorrefmark{6}VPIphotonics GmbH, Germany.
Email: shi.li@vpiphotonics.com}
\IEEEauthorblockA{\IEEEauthorrefmark{7}TelemaxX Telekommunikation GmbH, Germany.
Email: schweikert@telemaxx.de}
}

\maketitle

\newcommand{\figref}[1]{Fig.~\ref{#1}}
\newcommand{\secref}[1]{Sec.~\ref{#1}}
\newcommand{\secrefI}{\secref{sec:introduction}}
\newcommand{\secrefII}{\secref{sec:applications}}
\newcommand{\secrefIII}{\secref{sec:backbone}}
\newcommand{\secrefIV}{\secref{sec:ai}}

\begin{abstract}
Smart cities will be characterized by a variety of intelligent and networked services, each with specific requirements for the underlying network infrastructure. While smart city architectures and services have been studied extensively, little attention has been paid to the network technology. The \kiglis research project, consisting of a consortium of companies, universities and research institutions, focuses on artificial intelligence for optimizing fiber-optic networks of a smart city, with a special focus on future mobility applications, such as automated driving.
In this paper, we present early results on our process of collecting smart city requirements for communication networks, which will lead towards reference infrastructure and architecture solutions. Finally, we suggest directions in which artificial intelligence will improve smart city networks.
\end{abstract}

\begin{IEEEkeywords}
artificial intelligence, smart city, network infrastructure, fiber optics, automated driving
\end{IEEEkeywords}

\section{Introduction}
\label{sec:introduction}
A smart city can be defined as an urban area that uses information and communication technologies to become more sustainable, drive economic growth, increase citizen welfare, and improve the quality of life. It is characterized by a high degree of digitization and based upon a variety of interconnected services in the fields of economy, mobility, living, governing, and social affairs \cite{giffinger_smart_2007}. 
However, many technical hurdles need to be overcome before such a smart environment becomes a reality. In particular, the services differ strongly with respect to the requirements imposed on the network infrastructure. For instance,
low latency and high reliability are essential for real-time applications, whereas other services are not time-critical, but may produce a large amount of data that needs to be transmitted and processed. In addition, the heterogeneity of services can complicate interoperability. There can also be legal constraints, such as the German draft bill on automated driving \cite{bundesregierungEntwurfGesetzesZur2021}, which in its current form requires such vehicles to have a permanent network connection to allow a technical supervisor to remotely monitor and potentially assist the automated systems.

Operation of such services requires a network infrastructure with an intelligent and flexible management. Such a network needs to provide wireless connectivity to a massive number of endpoints, e.g. for car-to-network communication. A connection to the optical fiber backbone and to data centers must be guaranteed to handle high data volumes. This can be achieved by fixed-mobile convergence and a technology mix consisting of wireless access networks, long range wide area networks (LoRa-WANs), passive optical networks (PONs), and point-to-point (P2P) fiber solutions. Early concepts for this topic resulted from the \emph{Connected OFCity Challenge} \cite{Tucker:17}. However, the challenge did not take a closer look at operational details. \kiglis addresses the envisioned optical network solutions and how they can be enhanced by artificial intelligence (AI). \kiglis focuses on three key pillars: 1) AI for improving digital signal processing (DSP) in optical access networks, 2) AI for traffic management and resource allocations in such networks, and 3) AI for supporting optical network infrastructure planning. The aim is a final demonstration within the \emph{Test Area Autonomous Driving Baden-W\"urttemberg} in Karlsruhe, Germany \cite{Fleck:2018}.
\section{Service-driven network requirements}
\label{sec:applications}
Services of a smart city impose multidimensional demands on the network, which can vary immensely. At scale, some services need to be prioritized above others. We will give an overview about the variety of services in \secref{sec:usecases_broad} and introduce our analysis approach in \secref{sec:usecase_analysis}.

\subsection{Smart City Services}\label{sec:usecases_broad}
\kiglis focuses on services in the area of mobility solutions \cite{guerrero-ib_sensor_2018}. As an example, smart parking \cite{Klemm:2016} and smart waste management are already being applied in pilot projects \cite{noauthor_smart_nodate}. Visual surveillance \cite{kumaran_anomaly_2021} or audio event detectors \cite{almaadeed_automatic_2018} are proposed to detect road traffic anomalies, such as car accidents, and allow rapid intervention. Road safety can be further increased by digital traffic signs in vehicles \cite{bhatt_smart_nodate}, roadside unit (RSU) based driver warnings \cite{branquinho_efficient_2020}, or vehicle-to-pedestrian (V2P) communication \cite{peng_jing_car--pedestrian_2017}.

For steady traffic flows, congestion avoidance can be achieved with dynamic allocation of lanes to different types of transportation \cite{wang_dynamic_2016} or intersection management \cite{namazi_intelligent_2019}. Automated driving will contribute to both road safety and traffic flow \cite{talebpourInfluenceAutonomousConnected2015}.

\begin{figure}[htbp]
\centering
\resizebox{0.4\textwidth}{!}{

\definecolor{colorblind_one}{HTML}{648FFF}
\definecolor{colorblind_two}{HTML}{785EF0}
\definecolor{colorblind_three}{HTML}{DC267F}
\definecolor{colorblind_four}{HTML}{FE6100}

\begin{tikzpicture}

\footnotesize

\begin{axis}[xmode=log, ymode=log,
width=3in,
height=1.8in,
xlabel=Bandwidth (Gbit/s),
xtick={0.00001, 0.001, 0.1 ,10},
ylabel=Latency (s), 
ytick={10, 1, 0.1, 0.01, 0.001, 0.0001},
y dir=reverse, legend pos=south east, every node near coord/.append style={font=\footnotesize}]
    \addplot[
        scatter,only marks,mark size=3pt,scatter src=explicit symbolic,
        scatter/classes={
            a={mark=square*,colorblind_one},
            b={mark=triangle*,colorblind_two},
            c={mark=o,draw=colorblind_three,fill=colorblind_three},
            d={mark=x,draw=colorblind_four,fill=colorblind_four}
        },
        scatter src=explicit symbolic,
        nodes near coords*={\Label},
        visualization depends on={value \thisrow{label} \as \Label} 
    ]
    table[x=x,y=y,meta=class]{
        x    y    class label
        0.001 0.01 a CACC
        0.00008 0.0007 a ~
        0.001 0.001 a ~
        0.0003 0.01 a ~
        0.00005 0.01 a ~
        
        0.00008 5 b RSW
        0.000015 9 b ~
        0.0006 7 b ~
        0.0008 1 b ~
        0.00003 0.09 b ~
        0.007 0.2 b ~
        0.00004 1 b ~
        0.00008 1.2 b ~
        0.005 0.8 b ~
        0.00004 10 b ~

        0.7 0.1 c RA
        0.5 0.02 c ~
        2 0.003 c ~
        0.7 0.007 c ~
        0.08 0.06 c ~
        0.03 0.003 c ~
        0.3 0.003 c ~
        0.05 0.001 c ~

        1 1 d VS
        0.09 0.25 d ~
        3 0.5 d ~
        0.03 1 d ~
        5 2 d ~
        0.07 3 d ~
        7 5 d ~
    };
\end{axis}
\end{tikzpicture}}
\caption{Requirements of the analyzed services with representatives in each cluster: CACC, RA, VS, and RSW.}
\label{fig:requirements_applications_smart_city}
\end{figure}
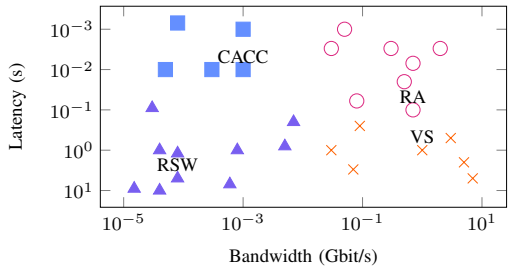

The benefits of smart city services increase with the level of interconnection. However, such communication and data processing require an appropriate network infrastructure and hence a detailed analysis of the specific network requirements of smart city services.

\subsection{Service Analysis}\label{sec:usecase_analysis}
\begin{figure}[htbp]
\centering
\resizebox{0.4\textwidth}{!}{\newcommand{\D}{6} 
\newcommand{\U}{3} 
\newcommand{\A}{360/\D} 

\definecolor{colorblind_one}{HTML}{648FFF}
\definecolor{colorblind_two}{HTML}{785EF0}
\definecolor{colorblind_three}{HTML}{DC267F}
\definecolor{colorblind_four}{HTML}{FE6100}

\newdimen\R 
\R=3.5cm 
\newdimen\L 
\L=4.5cm

\begin{tikzpicture}[scale=1]
\Large 
  \path (0:0cm) coordinate (O); 

  \foreach \X in {1,...,\D}{
    \draw[line width=1pt] (\X*\A:0) -- (\X*\A:\R);
  }

  \draw [color=colorblind_one,line width=4pt,opacity=1]
    (1*\A:0.1*\R/\U) --
    (2*\A:0.1*\R/\U) --
    (3*\A:0.9*\R/\U) --
    (4*\A:1.9*\R/\U) --
    (5*\A:3.0*\R/\U) --
    (6*\A:2.0*\R/\U) -- cycle;
        
  \draw [color=colorblind_two,line width=4pt,opacity=1]
    (1*\A:0.05*\R/\U) --
    (2*\A:0.005*\R/\U) --
    (3*\A:2.2*\R/\U) --
    (4*\A:3*\R/\U) --
    (5*\A:2*\R/\U) --
    (6*\A:0.05*\R/\U) -- cycle;
    
  \draw [color=colorblind_three,line width=4pt,opacity=1]
    (1*\A:1.2*\R/\U) --
    (2*\A:2.4*\R/\U) --
    (3*\A:2.4*\R/\U) --
    (4*\A:1*\R/\U) --
    (5*\A:3.0*\R/\U) --
    (6*\A:1.0*\R/\U) -- cycle;
    
  \draw [color=colorblind_four,line width=4pt,opacity=1]
    (1*\A:2.4*\R/\U) --
    (2*\A:3.0*\R/\U) --
    (3*\A:0.9*\R/\U) --
    (4*\A:0.2*\R/\U) --
    (5*\A:1.9*\R/\U) --
    (6*\A:1.9*\R/\U) -- cycle;
    
  \foreach \Y in {0,...,\U}{
    \draw[opacity=0.3,line width=1pt] circle (\Y*\R/\U);
  }
  \foreach \Y in {0,...,\U}{
    \foreach \X in {1,...,\D}{
      \path (\X*\A:\Y*\R/\U) coordinate (D\X-\Y);
      \fill (D\X-\Y) circle (2pt);
    }
  }
    
\node[right] at (D1-1) {\large 10};
\node[right] at (D1-2) {\large 100};
\node[right] at (D2-1) {\large 0.1};
\node[right] at (D2-2) {\large 0.5};
\node[right] at (D2-3) {\large 1};
\node[below] at (D3-1) {\large 99.99};
\node[above] at (D3-2) {\large 99.999};
\node[below] at (D3-3) {\large 99.9999};
\node[right] at (D4-1) {\large 50};
\node[right] at (D4-2) {\large 120};
\node[right] at (D4-3) {\large 250};
\node[right] at (D5-1) {\large 10};
\node[right] at (D5-2) {\large 1};
\node[right] at (D5-3) {\large 0.1};
\node[below] at (D6-1) {\large 100};
\node[below] at (D6-2) {\large 10};
\node[below] at (D6-1) {\large 1};

  \path (1*\A:\L+0.1cm) node[align=center] (L1) {Sustainable\\data rate (MBit/s)};
  \path (2*\A:\L+0.1cm) node[align=center] (L2) {Peak data rate\\(GBit/s)};
  \path (3*\A-0.1cm:\L+0.6cm) node (L3) {Reliability (\%)};
  \path (4*\A:\L) node[align=center] (L4) {Mobility (km/h)};
  \path (5*\A:\L) node (L5) {Positioning accuracy (m)};
  \path (6*\A+0.1cm:\L+0.5cm) node[align=center] (L6) {Latency (ms)};





\draw [color=colorblind_one,line width=4pt,opacity=1] (-7cm, 4.2cm) -- (-6.5cm, 4.2cm) node[right,opacity=1,black] {CACC};

\draw [color=colorblind_three,line width=4pt,opacity=1] (-7cm, 3.5cm) -- (-6.5cm, 3.5cm)
node[right,opacity=1,black] {RA};

\draw [color=colorblind_four,line width=4pt,opacity=1] (-7cm, 2.8cm) -- (-6.5cm, 2.8cm)
node[right,opacity=1,black] {VS};

\draw [color=colorblind_two,line width=4pt,opacity=1] (-7cm, 2.1cm) -- (-6.5cm, 2.1cm) node[right,opacity=1,black] {RSW};
\end{tikzpicture}}
\caption{Spider diagram for the multidimensional visualization of the requirements of the four representative services CACC, RA, VS, and RSW.}
\label{fig:spiderweb}
\end{figure}
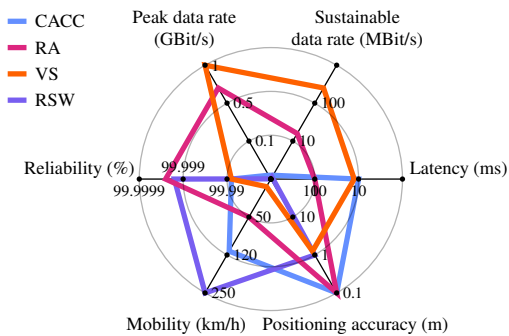

In \kiglis, 30 smart city services have been evaluated so far and subsequently assigned to four clusters depending on their bandwidth and latency requirements, as shown in \figref{fig:requirements_applications_smart_city}. From each cluster, one representative service has been chosen for further analysis, as shown in \figref{fig:spiderweb}. While high reliability, positioning accuracy, and latency are mandatory for Remote Assistance (RA) \cite{ackermanWhatFullAutonomy2021}, Video Surveillance (VS) \cite{alshammariIntelligentMultiCameraVideo2019} relies on high data rates. Cooperative Adaptive Cruise Control (CACC) \cite{wangReviewCooperativeAdaptive2018} and Road Safety Warnings (RSW) \cite{camaraPropagationPublicSafety2010} on the other hand have increased demand with respect to the velocity of the emitting devices.

In the following, particular emphasis is placed on RA, as it is a key technology for automated driving and one of the central scenarios considered in \kiglis. When automated driving systems identify corner-case or deadlock situations, human support can be requested. Following the SAE standard $J3016_{202104}$ \cite{saeJ3016CTaxonomyDefinitions2021}, automated driving features can only be supported by indirect \emph{remote assistance} \cite{ackermanWhatFullAutonomy2021} instead of direct \emph{remote driving} \cite{neumeierTeleoperationHolyGrail2019}. High bandwidth is required for the transmission of sensor data to the remote human agent. In contrast, latency requirements are relatively low since the automated driving system remains in control at all times. For precise assistance, demands on positioning accuracy and reliability are high \cite{3gpp_techreport_5g_v2x_services_2018}. Situations that require RA are most likely in urban environments where speeds of up to 50 km/h need to be tracked.

\setlength{\voffset}{.05in}

\begin{figure*}[htbp]
\centerline{\includegraphics[width=0.85\textwidth,
]{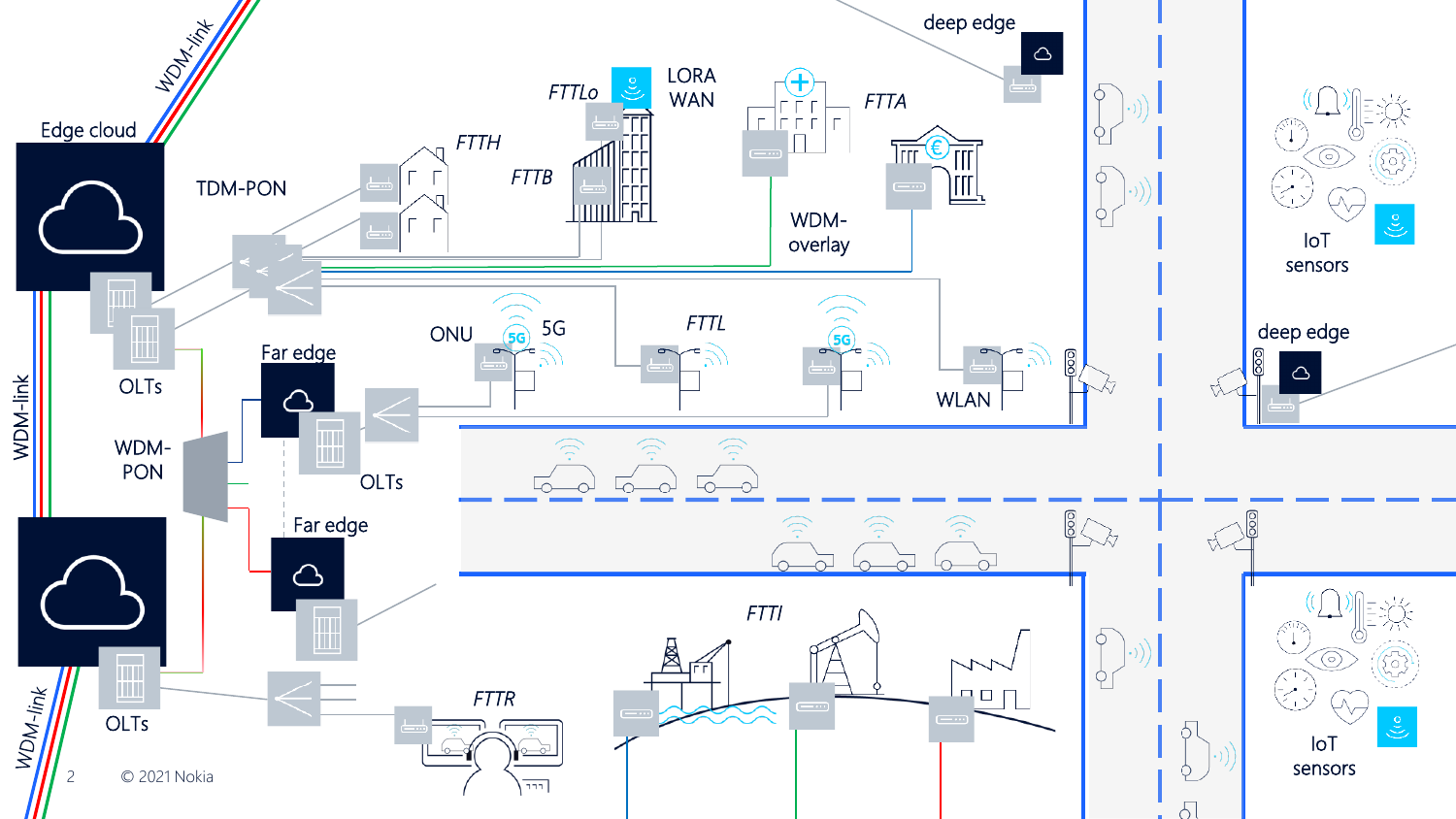}}
\caption{\kiglis smart city network infrastructure}
\label{fig:ref_infrastructure}
\end{figure*}

\section{Smart Architectures}
\label{sec:backbone}

Smart city services come with a diverse set of requirements that need to be satisfied to ensure high quality of service (QoS), see \figref{fig:spiderweb}. These must be considered with regard to networking properties, namely capacity, placement of network elements, fiber distances, redundancy, etc. Furthermore, the network has to ensure coexistence on legacy wireless and optical systems deployed.

A smart city network infrastructure developed within \kiglis is shown in \figref{fig:ref_infrastructure}, allowing to address the different services classified previously. Services with high latency and low bandwidth demands are well suited to be transported by LoRa\nobreakdash-WAN. Sensor data are aggregated in centralized LoRa\nobreakdash-WAN receiving units that are optically connected to the edge cloud data center.  
Public Wi\nobreakdash-Fi~6/7 offloads data from 5G radio access networks (RAN), e.g., for applications requiring moderate bandwidth and latency, while 5G RAN is used 
for enhanced mobile broadband and ultra reliable low latency communications. 

The content, processing and storage functionality as part of the data center cloud is distributed according to the bandwidth and latency needs of the services. Few large centralized edge clouds for moderate and low demanding services, a larger number of in-field far edge clouds for stringent services requirements, and a few service-dedicated deep edge cloud solutions, allowing for local processing and data offloading, are part of our smart city infrastructure. 

The backbone of the infrastructure is the optical fiber access network. In \kiglis, we envision the optical connectivity between edge clouds to be realized by P2P fibers using wavelength-division multiplexing (WDM) and high-capacity system solutions with 100~Gbit/s and beyond per wavelength. 

PONs, in turn, are used to bring fiber connectivity to households, business units, antenna sites, and towards city infrastructures.
 Using time-division multiplexing (TDM), optical line terminals (OLTs) located at the edge clouds will be connected  with the optical network units (ONUs) located deep in the network. Such TDM\nobreakdash-PONs combine the backhaul of the public Wi\nobreakdash-Fi~6/7 located at, e.g., lampposts with fiber-to-the-x\footnote{h = home, b = building, Lo = LoRa\nobreakdash-WAN, r = remote assistance, i = industry} solutions, see \figref{fig:ref_infrastructure}), thus, leveraging the synergy of deployment. Today's 10-gigabit capable XGS\nobreakdash-PON and also 25GS\nobreakdash-PON \cite{NextPON} can be envisioned to provide the connectivity over typical fiber distances of 10 to 20~km. Moreover, PONs allow for wavelength-overlays to connect local area and campus networks, e.g. of hospitals, to the data centers. 

Finally, to offer cost-efficient 5G RAN solutions for large scale deployment, virtualized RAN with cloudification of radio processing functions offers significant benefits by pooling of computing resources and simplifying antenna sites. \kiglis envisions the use of 5G small cell fronthaul transport links with capacity requirements dependent on the user data traffic and radio channel conditions. This enables the use of multiplexing technologies such as TDM\nobreakdash-PON \cite{9125631} for fronthauling, e.g.  25GS\nobreakdash-PON or future 50~Gbit/s TDM-PON \cite{9399710}. Fronthaul links, however, come with stringent latency requirements in the range of hundreds of $\mu$sec, which calls for locating the OLTs and most RAN entities into far edge clouds with less than 10~km fiber distance to small cell antenna sites. Far edge clouds, in turn, need to be optically connected with the centralized edge clouds. This can be done  via high-capacity TDM-PONs, e.g. 100~Gbit/s PONs \cite{9333413} or by using WDM-PONs with a large number of wavelengths (e.g. $>20$) and at least 25~Gbit/s per wavelength.

Core functionalities of all wireless solutions are centralized within the edge clouds. In our proposal, these functions are moved much deeper into the network compared to today's network infrastructures. This way the overall service latency can be reduced down to approximately 1~msec which will increase the QoS within the smart city.

\section{Artificial Intelligence for Smart Networks}
\label{sec:ai}
Operating a smart network requires managing and processing vast amounts of volatile data using forecasting, optimization, and reconfiguration methods. For this purpose, AI methods are identified, investigated, and demonstrated within \kiglis. The fundamental question of whether these methods outperform conventional tools not only in terms of performance, but also in terms of implementation complexity, latency, and reliability is to be clarified.
\paragraph*{AI for Fiber Network Signal Processing}

Using digital-analog converters in optical access networks enables the utilization of advanced DSP techniques. \cite{9333413}.
Here, machine learning might address technological challenges of a 100G-PON, e.g. compensating low-cost and low-bandwidth component distortions and channel impairments \cite{mlon}
or optimizing burst-mode reception, such as the alignment of the received signal powers in the upstream. In this context, neural networks (NNs) are ideally suited as signal equalizers. 
As an alternative to NNs, clustering algorithms offer the ability to detect point clouds in the constellation diagram, e.g. to support the phase synchronization in coherent optical systems. In addition to addressing signal impairments, the network service can be supplemented by classification techniques for forecasting the end of lifetime of employed hardware or the recognition of network anomalies, such as fiber defects.

\paragraph*{AI for Data Compression}
Beside DSP for the optical transmission channel, data compression is a central component for efficient data transmissions. 
There has been impressive progress in regards to the compression of image and speech data lately based on generative models \cite{ oquabLowBandwidthVideoChat2020}. 
Since these models are data-based and primarily explored in the field of image processing, multi-modal data streams of automated driving systems pose a particular challenge, as they must be correctly transmitted even in rare corner cases. 

\paragraph*{AI for Network Management}
While there is some early work on the incorporation of AI for a more efficient network operation \cite{aidba}, it is tailored for private customer access. However, due to the multitude of connected applications, network access can also be differentiated at the application level according to user classes and QoS. Hence, it is necessary to identify applications of AI and related methods ranging from classification to process mining and multi-criteria optimization on multiple levels, to optimize the network management. Here, \kiglis is focusing on the improvement of the dynamic bandwidth allocation, network automation for end-to-end slicing and computing resources along the edge-cloud continuum.

\paragraph*{AI for Fiber Network Infrastructure Planning}
Finally, AI techniques can help to reduce deployment costs and time for fiber-optic access networks. One major challenge when expanding fiber networks is the divergence and impreciseness of data on available infrastructure, piping, and cabling, leading to accidents and additional costs \cite{cost_ger}. \kiglis aims at using AI for merging and extracting data from geographic information systems, analog maps, and ground-penetrating radar images to synthesize a cartography of the urban environment.
These AI methods include image processing, semantic knowledge representation, as well as clustering.

\section{Summary \& Outlook}
\label{sec:outlook}

\kiglis investigates advantages and applicability of AI methods for realizing fixed-mobile converged networks in a smart city environment. The diverse requirements of services and the mixture of private and public interests push current network technologies to their limits. We have presented early results regarding the analysis of services as the basis for smart architectures with a focus on automated driving systems.
To overcome these challenges, we will identify, investigate and demonstrate the exploitation of suitable AI techniques for DSP in optical transmission links, for network traffic management and resource allocation, and for improving infrastructure planning processes. The shift of fiber access networks from private towards general public infrastructure poses multiple challenges and open questions that will be examined within \kiglis. For example, one challenge is how to utilize reserve capacities of existing, mainly residential, FTTx networks to support smart city endpoints.

While numerous publications will follow, we expect to perform a final demonstration within the \emph{Test Area Autonomous Driving Baden-W\"urttemberg} in Karlsruhe, Germany in 2023.
{\small
\bibliographystyle{IEEEtran}
\bibliography{references}
}

\end{document}